\documentclass[%
 aip,
 amsmath,amssymb,
 reprint,%
superscriptaddress,
]{revtex4-1}

\usepackage{graphicx}
\usepackage{dcolumn}
\usepackage{bm}

\usepackage[utf8]{inputenc}
\usepackage[T1]{fontenc}
\usepackage{mathptmx}

\begin{document}

\preprint{AIP/123-QED}

\title[Misinformation spreading on correlated multiplex networks]{Misinformation spreading on correlated multiplex networks}

\author{Jiajun Xian}
\affiliation{School of Computer Science and Engineering,
	University of Electronic Science and Technology of China, Chengdu, 611731, China}
\author{Dan Yang}
\affiliation{Web Sciences Center, School of Computer Science and Engineering,
	University of Electronic Science and Technology of China, Chengdu, 611731, China}
\author{Liming Pan}
\affiliation{Web Sciences Center, School of Computer Science and Engineering,
	University of Electronic Science and Technology of China, Chengdu, 611731, China}
\author{Wei Wang}
\email{wwzqbx@hotmail.com}
\affiliation{Cybersecurity Research Institute, Sichuan University, Chengdu 610065, China}
\affiliation{Web Sciences Center, School of Computer Science and Engineering,
	University of Electronic Science and Technology of China, Chengdu, 611731, China}
\email{wwzqbx@hotmail.com.}
\author{Zhen Wang}
\affiliation{School of Mechanical Engineering, Northwestern Polytechnical University, Xi’an 710072,China}
\affiliation{Center for OPTical IMagery Analysis and Learning, Northwestern Polytechnical University, Xi’an 710072, China.}

\date{\today}

\begin{abstract}
The numerous expanding online social networks offer fast channels for misinformation spreading, which could have a serious impact on socioeconomic systems. Researchers across multiple areas have paid attention to this issue with a view of addressing it. However, no systematical theoretical study has been performed to date on observing misinformation spreading on correlated multiplex networks. In this study, we propose a multiplex network-based misinformation spreading model, considering the fact that each individual can obtain misinformation from multiple platforms. Subsequently, we develop a heterogeneous edge-base compartmental theory to comprehend the spreading dynamics of our proposed model. In addition, we establish an analytical method based on stability analysis to obtain the misinformation outbreak threshold. On the basis of these theories, we finally analyze the influence of different dynamical and structural parameters on the misinformation spreading dynamics. Results show that the misinformation outbreak size $R(\infty)$ grows continuously with the effective transmission probability $\beta$ once $\beta$ exceeds a certain value, that is, the outbreak threshold $\beta_c$. A large average degrees, strong degree heterogeneity, or positive inter-layer correlation will reduce $\beta_c$, accelerating the outbreak of misinformation. Besides, increasing the degree heterogeneity or a more positive inter-layer correlation will both enlarge (reduce) $R(\infty)$ for small (large) values of $\beta$. Our systematic theoretical analysis results agree well with the numerical simulation results. Our proposed model and accurate theoretical analysis will serve as a useful framework to understand and predict the spreading dynamics of misinformation on multiplex networks, and thereby pave the way to address this serious issue.
\end{abstract}

\maketitle

\begin{quotation}
 Misinformation spreading have attracted substantial attention from multiple areas (e.g., network science, statistical physics, and computer science),  since the flooding of misinformation may result in serious impacts on socioeconomic systems. Researchers from the network science field have extended various classical network-based spreading models to study this issue by abstracting social and communication platforms into complex networks. In real life, each individual may be active on multiple platforms, and thus, it is necessary to consider spreading models with multiplex networks comprehensively. However, the theoretical research on misinformation spreading on correlated multiplex networks has not yet been conducted systematically. In this study, we contribute to this particular subject by proposing a multiplex network-based misinformation spreading model and developing a heterogeneous edge-based compartmental theory to describe the proposed model. Based on the developed theory, we further establish an analytical method to determine the misinformation outbreak threshold and analyze the influences of different dynamical and structural parameters on the spreading dynamics. Results reveals that misinformation is more likely to break out on multiplex networks with a larger average degree, stronger degree heterogeneity, or more positive inter-layer correlation. Besides, the misinformation outbreak size will be enlarged (or reduced) by an increase in the degree heterogeneity or additional positive inter-layer correlations for small (or large) values of the effective transmission probability. Our originally proposed model and precise theoretical analysis provide an in-depth understanding and accurate predictions of the underlying dynamics, which may stimulate further investigations.
\end{quotation}

\section{Introduction}
Misinformation is defined as false or inaccurate information such as false rumors, insults, hoaxes and spear phishing, which may arise from both deliberateness and carelessness~\cite{vosoughi2018spread,lazer2018science,
shao2018spread,kumar2018false,bovet2019influence}. The transmission efficiency of misinformation is tied up with developments in mass communication.
Various contemporary  online social networks and communication platforms (e.g., Twitter, Facebook, email) have provided
convenient channels for misinformation spreading.
As we know, misinformation can influence all aspects of an individual's life, including social relations, shopping behavior, political cognition, and financial security, to name just a few.
Fast and extensive spreading of misinformation could lead to negative impacts to socioeconomic systems by affecting individuals. Much attention has been geared toward tackling this issue from multiple fronts (e.g., network science, statistical physics, and computer science).
Understanding the mechanism behind misinformation spreading, building suitable spreading models to describe it, and combating it are the main focus areas~\cite{castellano2009statistical,
	pastor2015epidemic,wang2017unification}.

Researchers from the field of network science abstract social and communication platforms into complex networks, where nodes represent active users in these platforms and edges represent their relations~\cite{newman2010networks,jalili2017link}. By employing network representation, the spreading dynamics of the misinformation on these platforms can be studied using the existing theories and methods of network science. Many classic network-based spreading models can be extended to study misinformation spreading, such as rumor spreading models (e.g., Ignorant--Spreader--Recovered)~\cite{karp2000randomized,moreno2004dynamics,zhao2012sihr,zhou2007influence}, disease spreading models (e.g. SIR)~\cite{moreno2002epidemic,may2001infection,gross2006epidemic,
parshani2010epidemic}, and complex contagions (e.g., Watts threshold model and other models with social reinforcement effects)~\cite{watts2002simple,dodds2004universal,
wang2018effects,
zheng2013spreading}. In current literature, rumor spreading models are the most widely and directly used to describe misinformation spreading. The study of appropriate models can help us understand which types of the network topologies and spreading mechanisms facilitate the misinformation spreading and then develop effective strategies to combat it.

In real life, each individual may be active in multiple platforms; thus, comprehensively considering spreading models with multiplex networks is critical~\cite{wang2015evolutionary,boccaletti2014structure,kivela2014multilayer,wang2013interdependent,gao2012networks,pan2019optimal,wang2014degree}.
The strong dynamical correlations among the
states of neighbors and inextricably multiplex structures
could lead to difficulties in analytically studying the spreading models; thus, theoretical analysis becomes less accurate~\cite{de2018fundamentals}. There are two reasons for this. On the one hand, the characterization of the intricate network structure itself is a tough job. On the other hand, the dynamical correlations between neighboring nodes in the multiplex networks are also difficult to describe. Researchers have employed some theoretical methods to solve these problems, such as discrete  Markov chains~\cite{de2016physics,wang2013effect}, the percolation theory~\cite{dickison2012epidemics}, the heterogeneous mean field theory~\cite{saumell2012epidemic}, and the homogeneous edge-based compartmental theory~\cite{wang2018socialpre}.
Propagation research on multiplex networks has made good progress by benefiting from these applicable
theories ~\cite{de2018fundamentals,de2016physics}. Previous studies have revealed that multiplex networks can promote the spreading prevalence or lead to localization phenomenon in epidemic spreading and social contagions. In addition, Wang et al. observed that inter-layer degree correlations can inhibit spreading~\cite{wang2014asymmetrically} and turn the growing spread pattern from continuous to discontinuous~\cite{wang2018social}. Nevertheless, despite the achievement, theoretical research on misinformation spreading on correlated multiplex networks is yet to be performed systematically.

Given the current research status, we contribute to the study of misinformation spreading on correlated multiplex networks herein.
First, we propose a multiplex network-based misinformation spreading model, wherein the multiplex networks are considered to be a duplex system with two layers (also called as sub-networks). Specifically, we consider the spreading dynamics of misinformation on two-layer coupled networks, corresponding to the fact that each individual can obtain misinformation from multiple platforms. Subsequently, we develop a heterogeneous edge-based compartmental theory to comprehend the spreading dynamics of the proposed model. In addition, we establish an analytical method based on stability analysis to obtain the misinformation outbreak threshold. Finally, we analyze the influence of different dynamical and structural parameters on the spreading dynamics of misinformation on multiplex networks.

The remainder of this paper is organized as follows. In Section \ref{sec:model} we describe the multiplex network-based misinformation spreading model in detail. In Section \ref{sec:theory}, we systematically introduce the theoretical analysis. In Section \ref{sec:simulation}, we present extensive numerical stochastic simulations with respect to the theoretical predictions. In Section \ref{sec:conclusion}, we provide the conclusions.

\section{Model description} \label{sec:model}

In this section we will first introduce multiplex networks and their inter-layer correlations and then describe the spreading model of misinformation.

\subsection{Multiplex networks with inter-layer correlations}\label{sec:multi-net}

As illustrated in Fig. 1(b), we consider a duplex system with two layers/subnetworks, denoted as $1$ and $2$.
Each layer has $N$ nodes, which represent $N$ individuals, and the edges among these nodes represent the social connections among these individuals. Different layers could have distinct inner-layer structures and therefore dynamical properties. Nodes in the two layers are matched one-to-one, meaning that one individual has two distinct channels to spread misinformation. To be specific, for an individual denoted by $u$, there should be a pair of replica nodes $u_{1}$ and $u_{2}$ representing him/her in the corresponding layer labeled by their subscripts.
Let $k_1 (k_2)$ be the inner-layer degree of node $u_1 (u_2)$ in layer $1 (2)$. The degree distribution of the multiplex networks is then given by the joint distribution $p(k_{1},k_{2})$, which stands for the probability that a random selected node pair $(u_1,u_2)$ has an inner-layer degree $k_1$ in layer $1$ and $k_2$ in layer $2$.

Initially, we use a generalized configuration model~\cite{catanzaro2005generation} to generate multiplex network based on the join degree distribution $p(k_{1},k_{2})$. The main generation steps are as follows: (1) generate a degree pair sequence following the joint distribution $p(k_{1},k_{2})$; (2) assign a total number of $k^{i}_{1}$ and $k^{i}_{2}$ edge stubs to each node pair $u^{i}_{1}$ and $u^{i}_{2}$; (3) randomly select two stubs from the same layer to create an edge; and (4) repeat process (3) until there are no stubs left. Self-loops and multiple edges are prohibited throughout the process.
The configuration model stipulates that inner-layer degree correlations are negligible for large and sparse sub-networks. However inter-layer degree correlations can exist and will be captured by the joint distribution $p(k_{1},k_{2})$. Given the inadequacy of Pearson’s coefficient in measuring the correlation of heterogeneous sequences~\cite{yang2017lower,guo2017bounds}, we employ the Spearman rank correlation coefficient~\cite{lee2012correlated,cho2010correlated,wang2014asymmetrically} to quantify inter-layer correlation, which is defined as
\begin{equation}\label{eq:xiSIR}
m_s=1-6\frac{\sum_{i=1}^Nd_i^2}{N(N^2-1)},
\end{equation}
where $d_i$ is the difference between degree rankings of node pair $u^{i}_{1}$ and $u^{i}_{2}$ in the layer they belong to. The value of $m_s$ ranges from $-1$ to $1$, where values above $0$ indicate positive correlations and values below $0$ indicate negative correlations. The larger the absolute value, the stronger the positive or negative correlation. The role of inter-layer correlations in the spreading dynamics of misinformation on multiplex networks is discussed in Sec. \ref{sec:simulation}.

\subsection{Misinformation spreading model}\label{sec:msm}
\begin{figure}
	\centering
	\includegraphics[width=0.44\textwidth]{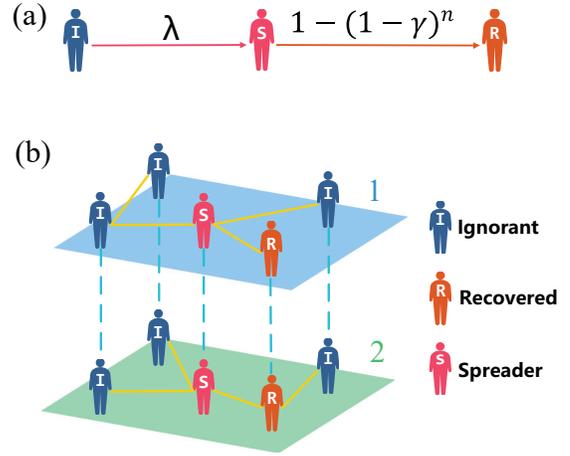}
	\caption{(Color online) Illustrations of (a) the generalized ignorant-spreader-recovered (SIR) model, and (b) the misinformation spreading model on multiplex networks. In (a), $\lambda$ denotes the transmission probability when individuals in ignorant states get contacted by spreaders, ${n}$ represents the total number of neighbors in the spreader or recovered states that the spreader individual has in both layers, and $\gamma$ is the unit recovery probability wherein the spreader individual has only one neighbor in the spreader or recovered state. The nodes in the two layers of the multiplex networks in (b) are matched one-to-one, and thus always have identical states.}\label{multiNet}
\end{figure}

We use a generalized ignorant-spreader-recovered (ISR) model \cite{catanzaro2005generation} to describe the dynamics of misinformation spreading on multiplex networks. Nodes in the networks transit in three possible states, that is, (1) ignorant state (I), in which the node has not known the misinformation; (2) spreader state (S), where
the node is willing to transmit the misinformation to his neighbors; and (3)
recovered state (R), in which the node has lost interest in the
misinformation.
Fig. \ref{multiNet}(a) illustrates the manner in which the nodes transit between states. For every time step, each spreader node in layer $a\in\{1,2\}$ first contacts all his/her neighboring ignorant nodes, and each contacted neighbor becomes a new spreader at the next time step with a probability of $\lambda$.
Then, each node in the spreader state turns into the recovered state with a probability of $1-(1-\gamma)^{n}$, where
${n}$ denotes the total number of neighbors in the spreader or recovered states that the spreader individual has in both two layers, and $\gamma$ denotes the unit recover probability that the spreader individual has only one neighbor in spreader or recovered state.
Once an individual in layer $a\in\{1,2\}$ moves into the spreader or recovered state, his/her counterpart also gets into the spreader or recovered state at the same time step, as they correspond to the same individual. Finally, all spreaders may lose interest in the misinformation and the dynamics will be terminated once there are no nodes in the spreader state.

\begin{figure*}
	\centering
	\includegraphics[width=0.9\linewidth]{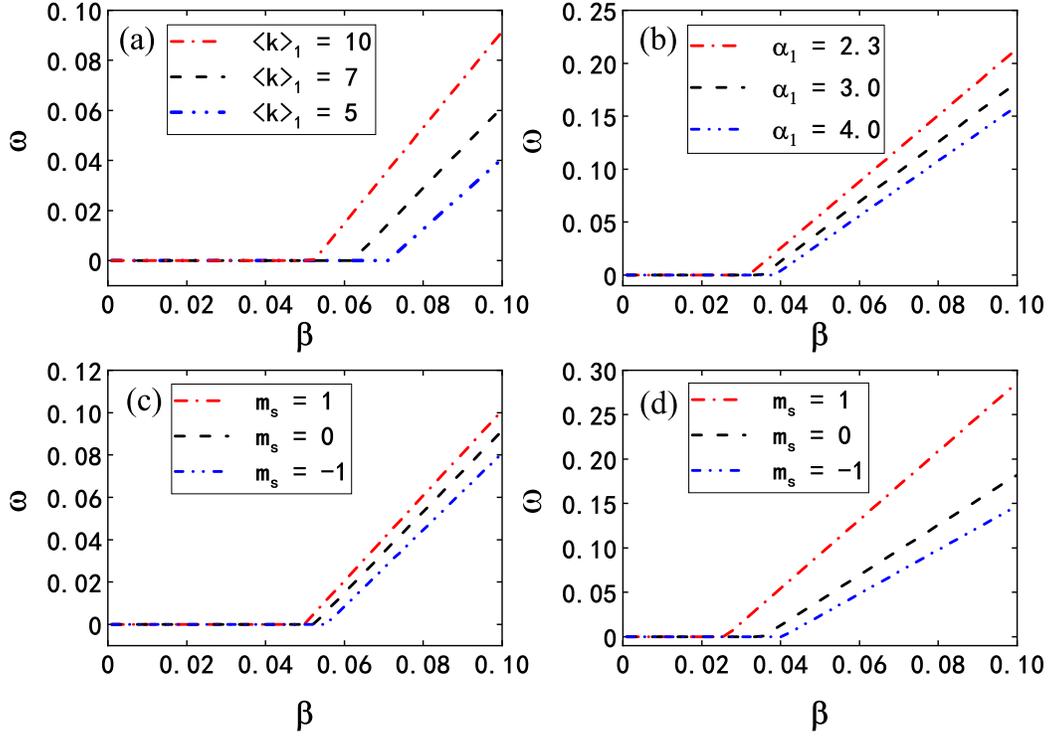}
	\caption{(Color online) Leading eigenvalue $\omega$ of the Jacobian matrix versus $\beta$ for typical networks, that is, (a) ER--ER multiplex networks with different average degrees $k_{1}$ as labeled in the subplot, set $k_{2}=10$; (b) SF--SF multiplex networks with different degree exponents $\alpha_{1}$ as labeled in the subplot, set $\alpha_{2}=3.0$; (c) ER--ER multiplex networks of average degrees $\langle k\rangle_1=\langle k\rangle_2=10$ with different inter-layer correlations, as labeled in the subplot; and (d) SF--SF multiplex networks of degree exponents $ \alpha_{1}=\alpha_{2}=3.0 $ with different inter-layer correlations, as labeled in the subplot. For SF--SF multiplex networks, the average degrees are all set to $10$, and the maximal degrees are all set to $\sqrt N$.  }\label{thredline}
\end{figure*}
\begin{figure*}
	\centering
	\includegraphics[width=0.9\textwidth]{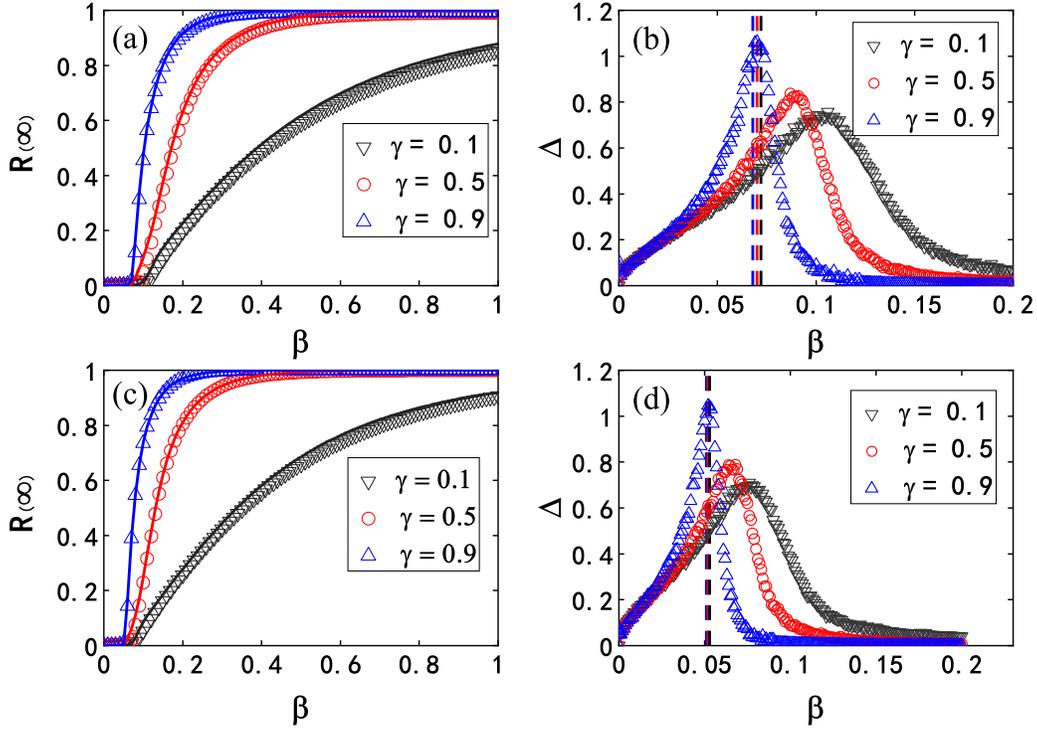}
	\caption{(Color online) Dynamics of spreading of misinformation on uncorrelated ER-ER multiplex networks. The final misinformation outbreak size $R(\infty)$ of the uncorrelated multiplex networks of average degrees (a) $\left\langle k\right\rangle_{1}=10$,$\left\langle k\right\rangle_{2}=5$ and (c) $\left\langle k\right\rangle_{1}=\left\langle k\right\rangle_{2}=10$, versus effective transmission probability $\beta$, with different unit recovery rate $\gamma$, as labeled in the subplot. The variability $\Delta$ of $R(\infty)$ on the uncorrelated multiplex networks of average degrees (b) $\left\langle k\right\rangle_{1}=10$,$\left\langle k\right\rangle_{2}=5$ and (d) $\left\langle k\right\rangle_{1}=\left\langle k\right\rangle_{2}=10$, versus $\beta$, with a different unit recovery rate $\gamma$, as labeled in the subplot. The dotted lines in (b) and (c) mark the positions of the theoretical threshold obtained in Sec. 3.2. The variability $\Delta$ is obtained by performing at least $2000$ independent simulations on a fixed multiplex network with a set $\gamma$ for each value of $\beta$. }\label{erRR}
\end{figure*}

\begin{figure}
	\centering
	\includegraphics[width=0.48\textwidth]{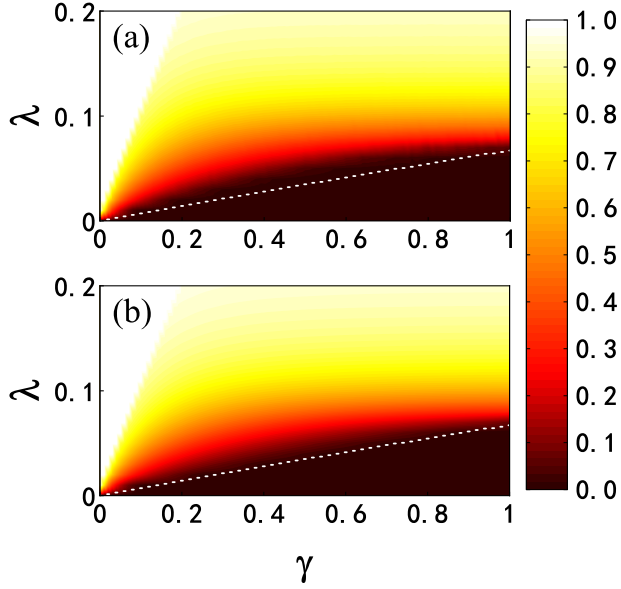}
	\caption{(Color online) (a) Simulation results and (b) theoretical predictions of $R(\infty)$ versus $\lambda$ and $\gamma$ on the uncorrelated ER-ER multiplex networks with $\left\langle k\right\rangle_{1}=10$ and $\left\langle k\right\rangle_{2}=5$. The white dotted lines mark the positions of the theoretical threshold obtained in Sec. 3.2. }\label{er3d}
\end{figure}

\begin{figure*}
	\centering
	\includegraphics[width=0.9\textwidth]{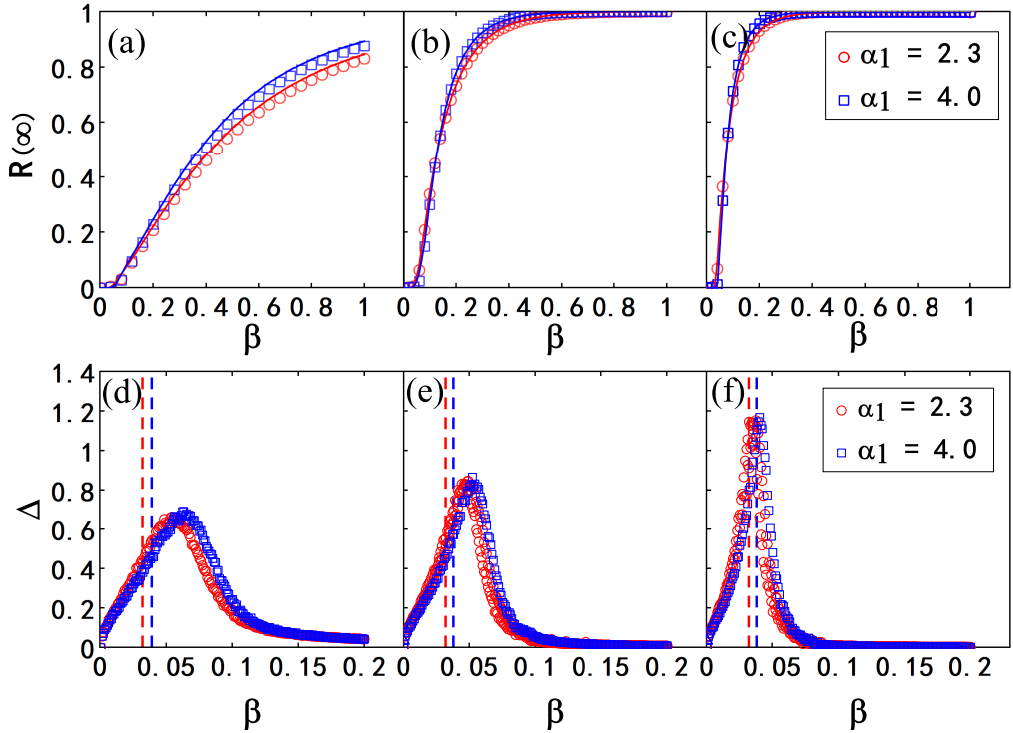}
	\caption{(Color online) Effects of degree heterogeneity on the spreading dynamics of misinformation on SF--SF multiplex networks. Misinformation outbreak size $R(\infty)$ versus $\beta$ on SF-SF multiplex networks when the unit recovery rate is set as (a) $\gamma=0.1$, (b)  $\gamma=0.5$, and (c) $\gamma=0.9$. Variability $\Delta$ of $R(\infty)$ when the unit recovery rate is set as (d) $\gamma=0.1$, (e) $\gamma=0.5$, and (f) $\gamma=0.9$. The exponents of the two groups of SF-SF multiplex networks are set as (1) $\alpha_{1}=2.3$ and $\alpha_{2}=3.0$ (corresponding results denoted by red circles), and (2) $\alpha_{1}=4.0$ and $\alpha_{2}=3.0$ (corresponding results denoted by blue squares), respectively.
		The dotted lines mark the positions of the corresponding theoretical threshold values obtained in Sec. 3.2.
	}\label{SFcorr}
\end{figure*}

\begin{figure*}
	\centering
	\includegraphics[width=0.9\textwidth]{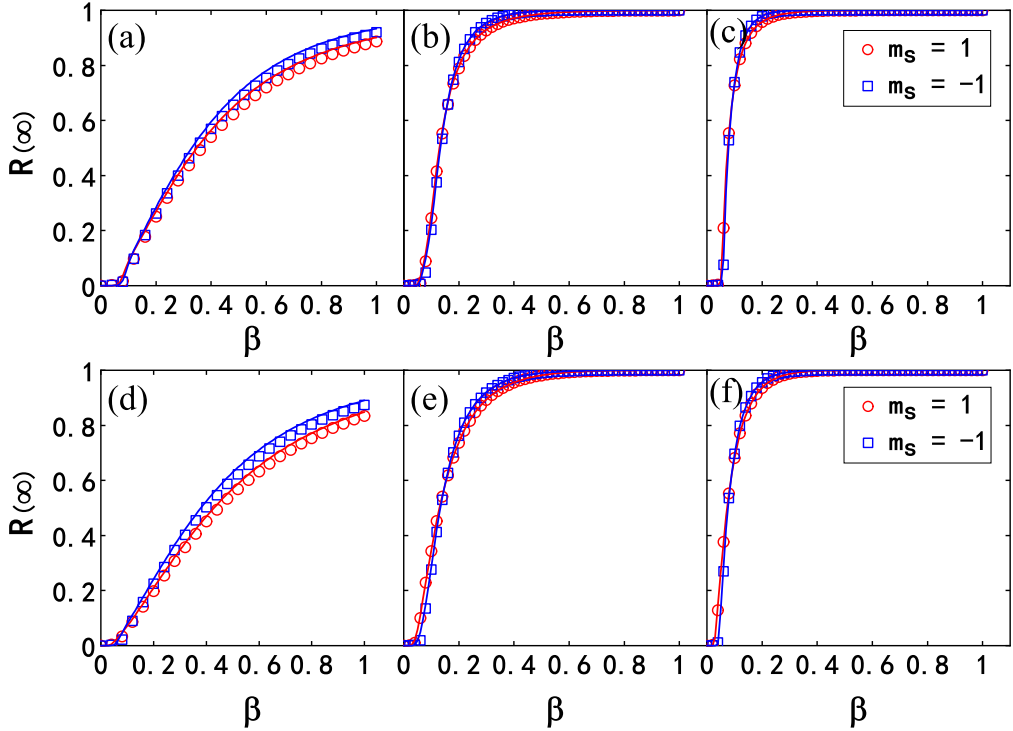}
	\caption{(Color online) Spreading dynamics of misinformation on multiplex networks with inter-layer correlations. Misinformation outbreak size $R(\infty)$ versus $\beta$ on ER-ER multiplex networks of different inter-layer correlations, with (a) $\gamma=0.1$, (b) $\gamma=0.5$, (c) $\gamma=0.9$. Misinformation outbreak size $R(\infty)$ versus $\beta$ on SF-SF multiplex networks of different inter-layer correlations with (d) $\gamma=0.1$, (e) $\gamma=0.5$, (f) $\gamma=0.9$. The average degree of layers 1 and 2 are set as $\left\langle k\right\rangle_{1}=\left\langle k\right\rangle_{2}=10$. For scale-free networks, we set the degree exponent as $\alpha_{1}=\alpha_{2}=3.0$. The red circle and blue squares denote the results of multiplex networks with inter-layer correlation $m_{s}=1$ and $m_{s}=-1$, respectively.
	}\label{corrRR}
\end{figure*}

\begin{figure*}
	\centering
	\includegraphics[width=0.9\textwidth]{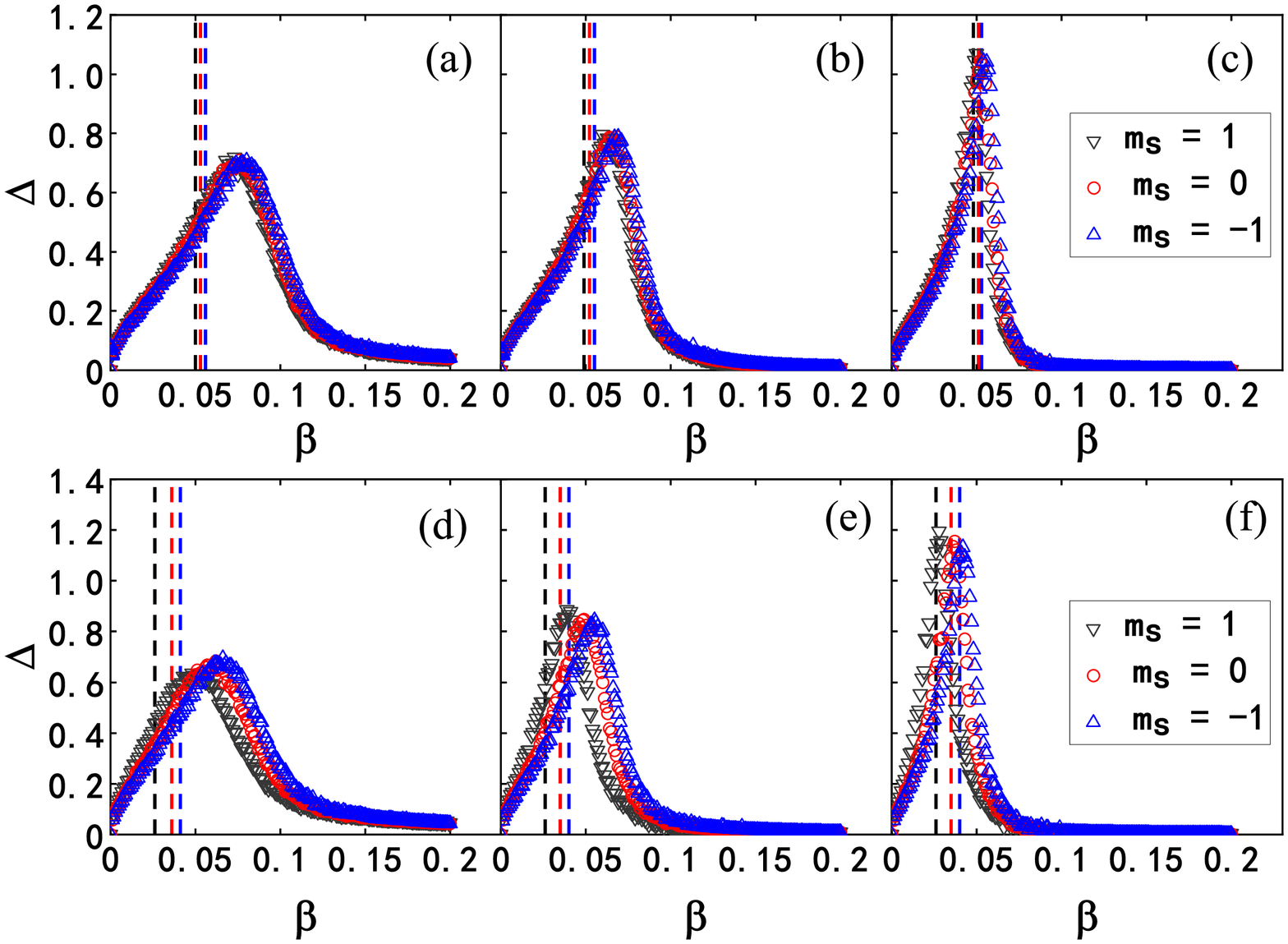}
	\caption{(Color online) Influence of inter-layer correlations on the outbreak threshold values. Variability $\Delta$ versus $\beta$ on ER-ER multiplex networks of different inter-layer correlations, with (a) $\gamma=0.1$, (b) $\gamma=0.5$, (c) $\gamma=0.9$. Variability $\Delta$ versus $\beta$ on SF-SF multiplex networks of different inter-layer correlations, with (a) $\gamma=0.1$, (b) $\gamma=0.5$, (c) $\gamma=0.9$. The average degree of layers 1 and 2 are set as $\left\langle k\right\rangle_{1}=\left\langle k\right\rangle_{2}=10$. For scale-free networks, we set the degree exponent as $\alpha_{1}=\alpha_{2}=3.0$. The black inverted triangles, red circle and blue regular triangle denote the results of multiplex networks with inter-layer correlation $m_{s}=1$, $m_{s}=0$ and $m_{s}=-1$, respectively.
	The dotted lines mark the positions of the corresponding theoretical threshold values obtained in Sec. 3.2.
	}\label{thredcorr}
\end{figure*}
\section{Theoretical analysis} \label{sec:theory}

According to the misinformation spreading model described in Section 2, the transition of each node’s state (i.e., the ignorant turns into spreader or the spreader becomes recovered) is strongly correlated to the state of his/her neighbors.
Thus, we develop the heterogeneous edge-based compartmental theory,
which is inspired by Refs.~\cite{miller2011edge,volz2008sir,miller2009spread,
valdez2013temporal,valdez2012temporal,valdez2012intermittent,wang2015dynamics}, to comprehend the misinformation spreading dynamics in the presence of strong dynamical correlations.
In this section, we first focus on the analysis of the misinformation outbreak size $R(\infty)$ versus effective transmission probability $\beta=\lambda/\gamma$. Then, we introduce an analytical method based on stability analysis to obtain the outbreak threshold $\beta_c$.

\subsection{Dynamics of misinformation spreading}

We denote $\theta_{a}(k_{a},k_{b},t)$ as the probability that node $v_{a}$ has not transmitted the misinformation to his/her neighboring node $u_{a}$ along a randomly selected edge up to time $t$ in layer $a\in \{ 1,2\} $, where $k_{a}$ and $k_{b}$ denote the degree of $v_{a}$ in layer $a$ and the degree of his/her counterpart in layer $b$, respectively, with $b\in \{1,2\}$ and $a\neq b$.
For simplicity, nodes with identical degrees are assumed to be the same with regard to their dynamical properties.
As node $v_{a}$ can be in three different states, we further divide $\theta_{a}(k_{a},k_{b},t)$ as
\begin{equation}\label{eq:xiSIR}
\theta_{a}(k_{a},k_{b},t)=\xi_{a}^{I}(k_{a},k_{b},t)+\xi_{a}^{S}(k_{a},k_{b},t)+\xi_{a}^{R}(k_{a},k_{b},t),
\end{equation}
where the value of $\xi_{a}^{I}(k_{a},k_{b},t)$, $\xi_{a}^{S}(k_{a},k_{b},t)$, and $\xi_{a}^{R}(k_{a},k_{b},t)$ represents
the probability of that node $v_{a}$ being in the ignorant, spreader, and recovered state, respectively, in layer $a$,
and has not transmitted misinformation to his/her neighboring node $u_{a}$ through a randomly selected edge by time $t$.

Inspired by the cavity theory, we assume that node $u_{a}$ is in the cavity state
(i.e., node $u_{a}$ cannot transmit misinformation to his/her neighbors but
can receive misinformation from his/her neighbors). With this assumption, one can move forward to capture some dynamical correlations among the states of nodes.
At each time step, the probability that node $v_{a}$ transmits the misinformation to node $u_{a}$ along a randomly selected edge
is equal to the probability that node $v_{a}$ is a spreader and transmits the misinformation to node $u_{a}$ at the exact time step in layer $a$.
Let $\mathcal{K}$ be the support of the joint degree distribution, that is, the values of degree pairs $\vec{k}=(k_{a},k_{b})$ that have non-vanishing probability. When considering a given realization of degree pair sequence, we also use $\mathcal{K}$ to denote the set of degree pairs appearing in the sequence when it causes no confusion.
Thus, the evolution of $\theta_{a}(k_{a},k_{b},t)$ for $\vec{k}\in \mathcal{K}$ is
\begin{equation}\label{eq:xiS}
\frac{d\theta_{a}(k_{a},k_{b},t)}{dt}=-\lambda\xi_{a}^{S}(k_{a},k_{b},t).
\end{equation}
Considering all possible degrees of node $v_{a}$, the average probability that node
$u_a$ has not received the misinformation from an edge by time $t$ is:
\begin{equation}\label{eq:theta}
\theta_{a}(t)=\frac{1}{\langle k \rangle_a}\sum_{\vec{k}\in \mathcal{K}}k_{a} p(k_{a},k_{b})\theta_{a}(k_{a},k_{b},t),
\end{equation}
where $\langle \cdot \rangle_a$ denotes the average degree of subnetwork $a$.
In layer $a$, an initial ignorant node $v_{a}$ can only get the misinformation from his/her other $(k_{a}-1)$ neighbors, as node $u_{a}$ is in the cavity state. Besides, nodes in ignorant state cannot transmit misinformation to neighbors; thus, $\xi_{a}^{I}(k_{a},k_{b},t)$
is equal to the probability of node $v_{a}$ in the ignorant state, that is,
\begin{equation}\label{eq:xiI}
\xi_{a}^{I}(k_{a},k_{b},t)=\theta_{a}(t)^{k_{a}-1}\theta_{b}(t)^{k_{b}}.
\end{equation}

Inferring from the misinformation spreading process described in Sec. \ref{sec:msm}, the
growth of $\xi_{a}^{R}(k_{a},k_{b},t)$ should simultaneously satisfy the following conditions: (1) the
spreader node $v_{a}$ does not transmit the misinformation to $u_{a}$ through the edge between them with a probability of $(1-\lambda)$, and (2) $v_{a}$ moves into the recovered state. For the second condition, we should count
the average numbers of neighbors in the spreader and recovered states that the spreader individual, represented by $v_{a}$, has in both layers.
To begin with, we obtain the probability that an edge connects to an ignorant node in layer $a$ as
\begin{equation}\label{eq:rhoI}
\rho_{a}(I)=\frac{1}{\langle k\rangle_a}\sum_{\vec{k}\in\mathcal{K}}k p(k_{a},k_{b})\theta_{a}(t)^{k_{a}}\theta_{b}(t)^{k_{b}}.
\end{equation}
Then, the probability that an edge connects to the spreader or recovered nodes in layer $a$ should be $\varphi _{a}=1-\rho_{a}(I)$, as the node can only exist in one of the three states. Follow the same steps, $\varphi _{b}$ can be obtained.
Eventually, assuming $u_{a}$ to be in the cavity state, we can obtain the average number of neighbors mentioned above as follows:
\begin{eqnarray}\label{eq:neighbor}
n_{a}(k_{a},k_{b},t)&=&1+r_{a}[(k_{a}-2)\varphi _{a}+k_{b}\varphi _{b}]\nonumber \\
					& &+r_{b}[(k_{b}-1)\varphi _{b}+(k_{a}-1)\varphi _{a}]
\end{eqnarray}
where
\begin{equation}
r_{a}^{a}=\frac{(k_{a}-1)\varphi _{a}}{(k_{a}-1)\varphi _{a}+k_{b}\varphi _{b}}
\end{equation}
and
\begin{equation}
r_{a}^{b}=\frac{k_{b}\varphi _{b}}{(k_{a}-1)\varphi _{a}+k_{b}\varphi _{b}}.
\end{equation}
In Eq. (\ref{eq:neighbor}), the first term within the braces on the right-hand side represents that the individual must connect to the neighbor who transmitted him/her the misinformation, which could be in either layer $a$ or $b$. If he/she gets the misinformation from layer $a$, then the average number of his/her spreader or recovered neighbors in layer $a$ and $b$ should be $(k_{a}-2)\left(1-\rho_{a}(I)\right)$ and $k_{b}(1-\rho_{b}(I))$, respectively. On the contrary, when he/she gets the misinformation from neighbor in layer $b$, the average number of his/her spreader or recovered neighbors in layer $a$ and $b$ should be $(k_{a}-1)\left(1-\rho_{a}(I)\right)$ and $(k_{b}-1)(1-\rho_{b}(I))$, respectively. To evaluate the net contributions from the two situations, we introduce $r_{a}^{a}$ and $r_{a}^{b}$, of which the values reflect the probability that the individual gets the misinformation whether from layer $a$ or $b$. The set-up is under the assumption that the layer with more non-I neighbors has greater opportunity to transmit the misinformation to the individual.
Now, node $v_{a}$ recovers with probability
\begin{equation}\label{eq:muk}
\mu_{a}(k_{a},k_{b},t)=1-(1-\gamma)^{n_{a}(k_{a},k_{b},t)},
\end{equation}
and the evolution of $\xi_{a}^{R}(k_{a},k_{b},t)$ should be
\begin{equation}\label{eq:xiR}
\frac{ d\xi^{R}_{a}(k_{a},k_{b},t)}{dt}=(1-\lambda)\xi^{S}_{a}(k_{a},k_{b},t)\mu_{a}(k_{a},k_{b},t).
\end{equation}
Combining Eqs. (\ref{eq:xiSIR}),(\ref{eq:xiS}), (\ref{eq:xiI}) and (\ref{eq:xiR}), we can obtain the value of $\theta_{a}(t)$ with $a\in \{1,2\}$.
Subsequently, we get the probability that a node pair $u_1$ and $u_2$ with degree pair $\vec{k}=(k_{1},k_{2})$ are ignorant by time $t$ as
\begin{equation}\label{eq:Ik}
I(\vec{k},t)=\theta_{1}(t)^{k_{1}}\theta_{2}(t)^{k_{2}}.
\end{equation}
Taking all possible values of $\vec{k}=(k_{1},k_{2})$ into
consideration, we can obtain the fraction (density) of ignorant
nodes at time $t$ as
\begin{equation}\label{eq:I}
I(t)=\sum_{\vec{k}\in\mathcal{K}}p(\vec{k})I(\vec{k},t).
\end{equation}
For the final state, that is, $t\rightarrow\infty$, the nodes in the multiplex networks can only be ignorant or recovered; thus, we obtain the misinformation outbreak size as follows
\begin{equation}\label{eq:I}
R(\infty)=1-I(\infty).
\end{equation}

\subsection{Threshold analysis}\label{sec:threshold}
Based on the dynamical equations obtained in the above subsection, we now calculate the spreading threshold.
There are four groups of dynamical variables $\theta_{a}(k_{a},k_{b},t)$, $\xi_{a}^{I}(k_{a},k_{b},t)$, $\xi_{a}^{S}(k_{a},k_{b},t)$, $\xi_{a}^{R}(k_{a},k_{b},t)$ for $a\in\{1,2\}$ and $k\in \mathcal{K}$. First note that $\xi_{a}^{I}(k_{a},k_{b},t)$ is fully determined by $\theta_{a}(k_{a},k_{b},t)$ and $\theta_{b}(k_{a},k_{b},t)$. Then, using the relation Eq. (\ref{eq:xiSIR}), $\xi_{a}^{S}(k_{a},k_{b},t)$ can be eliminated and the remaining equations are
\begin{equation}\label{eq:thresholdEquation}
\begin{large}
\left\{\begin{array}{l}\frac{ d\xi^{R}_{a}(k_{a},k_{b},t)}{dt}= g_{a}(k_{a},k_{b},t),\\
\\\frac{d\theta_{a}(k_{a},k_{b},t)}{dt}=f_{a}(k_{a},k_{b},t),\end{array}\right.
\end{large}
\end{equation}
with
\begin{equation}
f_{a}(k_{a},k_{b},t)=-\lambda \left(\theta_{a}(k_{a},k_{b},t)-\xi^{R}_{a}(k_{a},k_{b},t)-\xi^{I}_{a}(k_{a},k_{b},t)\right)
\end{equation}
and
\begin{equation}\label{}
g_{a}(k_{a},k_{b},t)=\left(1-\lambda^{-1}\right)\mu(k_{a},k_{b},t)f_a(k_{a},k_{b},t).
\end{equation}
There is always a trivial fixed point $\theta_{a}(k_{a},k_{b},t)=1$ and $\xi_{a}^{R}(k_{a},k_{b},t)=0$ for Eq.(\ref{eq:thresholdEquation}). The fixed point is not stable even below the spreading threshold, as any perturbation in the density of spreaders will lead to a non-vanishing number of recovered nodes in the stationary state.
However, below the spreading threshold, a small perturbation will result in $R\sim O(1/N)$, while above the spreading threshold, it will result in $R\sim O(1)$. Thus, below the spreading threshold, a small perturbation is expected to deviate the system from the unstable fixed point very slowly, while above the threshold, the deviation is much more significant. This is revealed locally in the eigenvalues of the Jacobian matrix of the system. When below the threshold, the leading eigenvalue $\omega$ of the Jacobian would be a small number very close to zero, while above the threshold, it deviates from zero and grows with $\beta$. To determine the spreading threshold, first, we compute the Jacobian matrix and check its leading eigenvalue at $\theta_{a}(k_{a},k_{b},t)=1$ and $\xi_{a}^{R}(k_{a},k_{b},t)=0$.
Let $C=\vert \mathcal{K}\vert$ be  the cardinality of $\mathcal{K}$. We define vectors $\Theta_a$ indexed by $\vec{k}\in\mathcal{K}$, with elements
$\Theta_a^{\vec{k}}=\theta_{a}(k_{a},k_{b},t)$ and $\Xi_a$ with
$\Xi_a^{\vec{k}}=\xi^{R}_{a}(k_{a},k_{b},t).$
Similarly, we define vectors $F_a$ and $G_a$ indexed by $\vec{k}\in\mathcal{K}$, with elements
$F_a^{\vec{k}}=f_{a}(k_{a},k_{b},t)$
and
$G_a^{\vec{k}}=g_{a}(k_{a},k_{b},t)$
respectively.
Then, the Jacobian matrix can be written in the following block matrix form:
\begin{equation}
J=
\begin{large}
\left(
\begin{array}{cccc}
\frac{\partial F_{a}}{\partial \Theta_{a}} &  \frac{\partial F_{a}}{\partial \Theta_{b}} & \frac{\partial F_{a}}{\partial \Xi_{a}} & \frac{\partial F_{a}}{\partial \Xi_{b}} \\\\
\frac{\partial F_{b}}{\partial \Theta_{a}} &  \frac{\partial F_{b}}{\partial \Theta_{b}} & \frac{\partial F_{b}}{\partial \Xi_{a}} & \frac{\partial F_{b}}{\partial \Xi_{b}} \\\\
\frac{\partial G_{a}}{\partial \Theta_{a}} &  \frac{\partial G_{a}}{\partial \Theta_{b}} & \frac{\partial G_{a}}{\partial \Xi_{a}} & \frac{\partial G_{a}}{\partial \Xi_{b}} \\\\
\frac{\partial G_{b}}{\partial \Theta_{a}} &  \frac{\partial G_{b}}{\partial \Theta_{b}} & \frac{\partial G_{b}}{\partial \Xi_{a}} & \frac{\partial G_{b}}{\partial \Xi_{b}} \\
\end{array}
\right),
\end{large}
\end{equation}
where the blocks correspond to contributions from different set of dynamical variables. For instance, $\frac{\partial F_{a}}{\partial \Theta_{a}}$ is a $C\times C$ matrix, where the entries are partial derivatives of $F_a$ to $\Theta_{a}$, which can be calculated as follows,
\begin{equation}\label{}
\frac{\partial f_{a}(k_{a},k_{b},t)}{\partial \theta_{a}(k_{a}^{\prime},k_{b}^{\prime},t)}
=\lambda\left(\frac{k_{a}^{\prime}(k_{a}-1) p(k_{a}^{\prime},k_{b}^{\prime})}{\langle k\rangle_a}-\delta_{\vec{k},\vec{k^{\prime}}}\right),
\end{equation}
where $\vec{k}=(k_a,k_b)$, $\vec{k^{\prime}}=(k_{a}^{\prime},k_{b}^{\prime})$
and
\begin{equation}\label{}
\delta_{\vec{k},\vec{k^{\prime}}}=
\left\{\begin{array}{lc}0,&\vec{k}\neq\vec{k^{\prime}},
\\1,&\vec{k}=\vec{k^{\prime}}.\end{array}\right.
\end{equation}
Similarly, we can get entries of all the other blocks.
Eventually, by substituting the general terms with the actual partial derivatives, we get the Jacobian matrix at the given point as:
\begin{equation}
J=
\begin{footnotesize}
\left(
\begin{array}{cccc}
\lambda (X^1-I)&\lambda Y^{12}&\lambda I&0\\
\lambda Y_{21}&\lambda (X^2-I)&0&\lambda I\\
\gamma\left(\lambda-1\right) (X^1-I)&\gamma\left(\lambda-1\right)Y^{12}& \gamma\left(\lambda-1\right)I&0\\
\gamma\left(\lambda-1\right)Y^{21}&\gamma\left(\lambda-1\right) (X^2-I)&0& \gamma\left(\lambda-1\right)I \\
\end{array}
\right),
\end{footnotesize}
\end{equation}
where $X^{a}$ is a $C\times C$ matrix  with elements
\begin{equation}
X^{\vec{k},\vec{k^{\prime}}}_a=\frac{k_{a}^{\prime}(k_{a}-1) p(k_{a}^{\prime},k_{b}^{\prime})}{\langle k\rangle_a},
\end{equation}
and $Y^{ab}$ is a $C\times C$ matrix  with elements
\begin{equation}
Y^{\vec{k},\vec{k^{\prime}}}_{ab}=\frac{k_{b}k_{b}^{\prime}p(k_{a}^{\prime},k_{b}^{\prime})}{\langle k\rangle_b}.
\end{equation}
As is shown in Fig. \ref{thredline}, we plot $\omega$ as a function of $\beta$ for typical multiplex networks. One can see that $\omega$ stays close to zero when $\beta$ is below a certain value $\beta_c$, which indicates that the fixed point $\theta_{a}(k_{a},k_{b},t)=1$ and $\xi_{a}^{R}(k_{a},k_{b},t)=0$ is shifted slightly by the perturbations. While $\beta$ exceeds $\beta_c$, $\omega$ deviates from zero significantly and grows with $\beta$. Thus, point $\beta_c$ can be understood as the spreading threshold. Our theoretical prediction shows that the threshold values can be influenced by the average degree, degree heterogeneity, and inter-layer correlations of the multiplex networks, which will be discussed in detail in the next section.
\section{Simulation results}\label{sec:simulation}
In this section, we perform extensive numerical simulations on artificial multiplex networks and compare the results with theoretical predictions.
Two typical multiplex networks are investigated, that is, ER-ER multiplex networks and SF-SF multiplex networks, of which each layer is a Erd\"{o}s-R\'{e}nyi (ER) network and scale-free (SF) network, respectively. SF networks display heterogeneous degree distributions, where the probability that a randomly selected node with degree $k$ follows a power-law distribution, is $p(k)\sim k^{-\alpha}$, where $\alpha$ denotes the degree exponent. Small degree exponents indicate high heterogeneity.
We first study the roles of dynamical parameters (e.g. $\lambda$ and $\gamma$) in the process of misinformation spreading, and then discuss the effects of different structural parameters (e.g. $\left\langle k\right\rangle$, $\alpha$, and $m_s$).
For all simulation results, we perform at least 100
independent realizations on a fixed multiplex network with $10000$ nodes in each layer to calculate the average values, which are further averaged over $50$ network realizations. The spreading process is initiated by randomly choosing five nodes as spreaders, while the other nodes are set to be ignorant.

There are two essential dynamical parameters in our proposed model, that is, $\lambda$ and $\gamma$, where $\lambda$ controls the transition of nodes from state I to S and $\gamma$ controls the recovery process.
To determine how these two parameters act in the spreading dynamics of misinformation, we consider the spreading on simple uncorrelated ER-ER multiplex networks with average degree of layers $1$ and $2$ as $\left\langle k\right\rangle_{1}=10$ and $\left\langle k\right\rangle_{2}=5$.
Fig. \ref{erRR}(a) demonstrates the misinformation outbreak size $R(\infty)$ versus the effective transmission probability $\beta$ on the underlined networks.
It can be seen that $R(\infty)$ stays close to zero before a certain value of $\beta$ is reached, which is understood as the outbreak threshold $\beta_c$.
Once $\beta$ exceeds $\beta_c$, the spreading breaks out and $R(\infty)$ grows with $\beta$.
Fig.~\ref{er3d}(a) [(b)] further shows the simulation results (theoretical
predictions) of $R(\infty)$ versus $\lambda$ and $\gamma$. The white dotted lines mark the positions of the theoretical thresholds obtained in Sec. 3.2, which divide the whole plane into two regions, that is, the misinformation outbreak region and the non-outbreak region.
To summarize, $\lambda$ and $\gamma$ jointly determine the spreading dynamics on the given multiplex networks, and misinformation outbreaks only when their ratio exceeds the threshold value. One can see that our theoretical predictions agree well with the simulations results.
We next consider porting the spreading process to networks with different
structures and compare the results to investigate the influence of different structural parameters.

Firstly, we focus on the effects of the average degree. In addition to the ER-ER multiplex networks mentioned in the above paragraph, we add another group of networks for comparison. In the additional group, we set $\left\langle k\right\rangle_{1}=\left\langle k\right\rangle_{2}=10$, keeping other structures the same with the original group.
Comparing the results of Fig. \ref{erRR}(a) and Fig. \ref{erRR}(c), we observe that with fixed $\lambda$ and $\gamma$, ER-ER multiplex networks with larger average degree will have bigger $R(\infty)$. However, the growing patterns of $R(\infty)$ remain the same with different average degrees. To obtain the outbreak threshold, we employ a numerical method mentioned in Ref.~\cite{shu2018social} by analyzing the peak of variability $\Delta$ of $R(\infty)$, where
\begin{equation}
\Delta=\frac{\sqrt{\left\langle R(\infty)^2\right\rangle-\left\langle R(\infty)\right\rangle^2}}{\left\langle R(\infty)\right\rangle}.
\end{equation}
The results of Fig. \ref{erRR}(b) and Fig. \ref{erRR}(d) demonstrate that an increase in the average degree will decrease the outbreak threshold. These results are comprehensible as a larger average degree indicates more connections between nodes, which gives more opportunity for each node to get access to misinformation. It can be seen that our theoretical predictions are in good agreement with the simulation results.

Secondly, we pay attention to the influence of the degree heterogeneity on the spreading dynamics, as most real-world networks display heterogeneous degree distribution.
Specifically, we study the spreading of misinformation on two groups of SF-SF multiplex networks with different degree exponents, that is, (a) $\alpha_{1}=2.3$, $\alpha_{2}=3.0$, and (b) $\alpha_{1}=4.0$, $\alpha_{2}=3.0$.
As shown in Figs. \ref{SFcorr}(a)-(c),
increasing the heterogeneity promotes (suppresses) the spreading of misinformation for small(large) value of $\beta$.
These results can be explained qualitatively by our theory.
From Eq. (\ref{eq:Ik}), we know that the state of each individual $i$ about the misinformation is determined jointly by the inner degree pair $(k_1^{i},k_2^{i})$ of his/her replica node pair. An increase of either $k_1^{i}$ or $k_2^{i}$ will allow the individual more probability to get into the spreader state. For each SF sub-network in the multiplex network, decreasing the degree exponents indicates that there are more nodes with large degrees in the network along with a large number of nodes with very small degrees. When $\beta$ is small, more hubs for smaller degree exponent will facilitate the spreading of misinformation, resulting in larger $R(\infty)$. However, when $\beta$ is large, the larger number of nodes with very small degrees in the heterogeneous networks adopt the misinformation with very small probability; thus, $R(\infty)$ becomes smaller.
It can also be observed from Figs.~\ref{SFcorr}(d)-(f) that strong heterogeneity will reduce the threshold value, for the existence of more hub nodes. Again, out theoretical predictions are verified by the simulation results.

Finally, we study the spreading dynamics of misinformation on multiplex networks with inter-layer correlations, to investigate the effects of $m_s$ on the spreading dynamics. Two groups of networks are employed here, that is, (a) ER-ER multiplex networks with different inter-layer correlations ($m_s=1$, $m_s=0$ and $m_s=-1$), and (b) SF-SF multiplex networks with different inter-layer correlations ($m_s=1$, $m_s=0$ and $m_s=-1$). The average degrees in all multiplex networks is set as $\left\langle k\right\rangle_{1}=\left\langle k\right\rangle_{2}=10$. For SF-SF multiplex networks, the degree exponents are set as $\alpha_{1}=\alpha_{2}=3.0$.
Fig. \ref{corrRR} reveals that, for small value of $\beta$, multiplex networks with positive inter-layer correlations (i.e., $m_{s}=1$) will have larger $R(\infty)$ than those with negative inter-layer degree correlations (i.e., $m_{s}=-1$). However, for large $\beta$, the opposite situation arises.
The results are consistent between the spreading of misinformation on ER-ER [see Figs. \ref{corrRR}(a)-(c)] and SF-SF [see
Figs.~\ref{corrRR}(d)-(f)] multiplex networks.
Our proposed theory can also give a qualitative explanation for these results.
As mentioned above, the probability of each individual $i$ to adopt the misinformation is determined by the inner degree pair $(k_1^{i},k_2^{i})$ of his/her replica node pair. We define those node pairs with both large (small) values of $(k_1^{i}$ and $k_2^{i})$ as hub-pairs (margin-pairs). From Eq. (\ref{eq:Ik}), we know that the hub-pairs are more likely to adopt the misinformation, while for the margin-pairs, the probability is small.
There will be more hub-pairs and margin-pairs in the multiplex networks with $m_s=1$ than those with $m_s=-1$. For small values of $\beta$, more hub-pairs will promote the spreading process, enlarging $R(\infty)$. However, for small values of $\beta$, the larger number of margin-pairs suppress the spreading, resulting in smaller $R(\infty)$. Fig. \ref{thredcorr} also demonstrates that
inter-layer correlations will influence the threshold values in both ER-ER [see Figs.~\ref{thredcorr}(a)-(c)] and SF-SF [see Figs.~\ref{thredcorr}(d)-(f)] multiplex networks.
The result shows that positive inter-layer correlations of multiplex networks will reduce the value of $\beta_c$. This stems from the existence of more hub-pairs in the multiplex networks with positive inter-layer correlations.
Note that our theoretical predictions of both $R(\infty)$ and $\beta_c$ coincide with the simulation results, regardless of the multiplex network structures.

\section{Conclusions}\label{sec:conclusion}
In consideration of the flooding misinformation in fast-growing online social networks, we carried out a systematic study on the spreading of misinformation on correlated multiplex networks. To be concrete, we originally proposed a correlated multiplex network-based model for the spreading of misinformation, and then comprehended the spreading dynamics by using a heterogeneous edge-based compartmental theory. Furthermore, we developed an analytical method based on stability analysis to obtain the outbreak threshold for spreading of misinformation. To the best of our knowledge, no previous studies have investigated this particular topic.

Through detailed theoretical analysis and simulation verification, in this study, we determined the influence of different dynamical and structural parameters on the spreading dynamics of misinformation on multiplex networks. The global misinformation outbreaks once the effective transmission probability exceeds the outbreak threshold, and the outbreak size grows with the effective transmission probability after the outbreak. Besides, the outbreak thresholds can be reduced under larger average degree, stronger degree heterogeneity and more positive inter-layer correlations. Moreover, higher degree heterogeneity and more positive inter-layer correlation will both promote (suppress) the spreading for small (large) value of the effective transmission probability. Our theoretical predictions agree well with the simulation results in all the cases studied.

Given the fact that misinformation flood on multiple online social platforms could lead to serious damages to socioeconomic systems, more attention and efforts are urgently needed to the study of spreading of misinformation.
We here present a useful framework based on an originally proposed model and precise theoretical analysis to systematically study the spreading dynamics of misinformation on correlated multiplex networks. This work offers deep understanding and accurate predictions of the underlying dynamics of spreading of misinformation, which will stimulate further works. For instance, our work will pave the way to designing effective combating strategies to address this issue. Additional studies on spreading of misinformation on multiplex networks with other abundant structures can also be carried out.

\begin{acknowledgments}
This work was partially supported by the China Postdoctoral Science Foundation (Grant No.~2018M631073), China Postdoctoral Science Special Foundation (Grant No.~2019T120829), National Natural Science Foundation of China (Grant No.~61603074), and Fundamental Research Funds for the Central Universities.
\end{acknowledgments}

\nocite{*}
\bibliography{xianjiajun}
\end{document}